# Some Remarks on the Model Theory of Epistemic Plausibility Models

Lorenz Demey*

Center for Logic and Analytic Philosophy
University of Leuven, Belgium
`lorenz.demey@hiw.kuleuven.be`

October 26, 2018

## 1 Introduction

Traditional epistemic logic can be seen as a particular branch of modal logic. Its semantics is defined in terms of Kripke models, and philosophical principles about knowledge (e.g. factivity: $K\varphi \to \varphi$) are shown to correspond to properties of the epistemic accessibility relation (e.g. reflexivity). By adding another (doxastic) accessibility relation, also *belief* can be treated in this framework. Belief is not assumed to be factive, but at least consistent ($\neg B\bot$), which corresponds to requiring the doxastic accessibility relation to be serial instead of reflexive. In this extended framework, one can study the interaction between knowledge and belief (e.g. is $K\varphi \to B\varphi$ a valid principle?); cf. [5, 8]. Furthermore, since this framework is still 'just' a (multi-)modal logic, it inherits the mathematically well-developed model theory of modal logic.

However, it is well-known that dynamic phenomena cannot be captured in this framework; cf. section 3.1 of [7]. To remedy this, epistemic plausibility models have been introduced (technical details will be presented later). In these models, one can again study knowledge, belief (and even other cognitive propositional attitudes), and their various interactions. Furthermore, this framework provides a realistic model of various dynamic phenomena, and thus solves the main problem of the previous approach. However, because epistemic plausibility models are much richer structures than Kripke models, they do not straightforwardly inherit the model-theoretical results of modal logic. Therefore, while epistemic plausibility structures are well-suited for modeling purposes, an extensive investigation of their model theory has been lacking so far.

The aim of the present paper is to fill exactly this gap, by initiating a systematic exploration of the model theory of epistemic plausibility models. Like in 'ordinary' modal logic, the focus will be on the notion of *bisimulation* — it turns out that finding the right generalization of this notion is not a trivial task. In Section 2, we introduce epistemic plausibility models and discuss some important operators which can be interpreted on such models, and their dynamic behaviour. In Section 3, we define various notions of bisimulations (parametrized by a language $\mathcal{L}$) and show that $\mathcal{L}$-bisimilarity implies $\mathcal{L}$-equivalence. We establish a Hennesy-Milner type theorem, and prove two undefinability results — thus shedding some light on the formal relationships between the various operators that can be interpreted on epistemic plausibility models. The notion of bisimulation for conditional belief, however, turns out to be unsatisfactory for several reasons. In Section 4, we discuss these reasons and explore two possible solutions: adding a modality to the language, and putting extra constraints on the models. In Section 5, we establish some results about the interaction between bisimulation and dynamic model changes.

---
*I wish to thank Johan van Benthem and Davide Grossi for their valuable comments on earlier versions of this paper. This research was funded by the University of Leuven's *Formal Epistemology Project*.

From a broader perspective, this paper can be seen as a reaction against a widespread trend in the technical modal logic literature (already since the 1960's), viz. the inclination to focus almost exclusively on the model theory of 'classical' single-quantifier modalities, while neglecting more-quantifier modalities (which are of central importance for applications in game theory, AI, philosophy, and linguistics).

Finally, it should be noted that this paper is mainly *exploratory* in nature. Especially with respect to the problem of finding bisimulations for conditional belief, the aim is to provide a (partial) map of the wide landscape of possible solutions, rather than to argue for the ultimate correctness of one of them.

## 2  Epistemic plausibility models

We now introduce epistemic plausibility models. Let $G$ be a non-empty set, whose elements will be called *agents*. Throughout this paper, we will keep the set of agents fixed, so that it can almost always be left implicit. Likewise, we assume that $Prop$ is a (countably infinite) set of proposition letters, which will also be kept fixed throughout the paper.

**Definition 1.** An *epistemic plausibility model* is a structure $\mathbb{M} = \langle W, \{\sim_i\}_{i \in G}, \{\leq_{i,w}\}_{i \in G}^{w \in W}, V \rangle$, where $W$ is a non-empty set of *states*, $\sim_i \subseteq W \times W$ is the *epistemic accessibility relation* for agent $i$, $\leq_{i,w} \subseteq W \times W$ is the *plausibility order* for agent $i$ at state $w$, and $V: Prop \to \wp(W)$ is a *valuation*.

As usual, $w \sim_i v$ is to be read as: "agent $i$ cannot epistemically distinguish between states $w$ and $v$". We assume this relation to be an equivalence relation. Furthermore, $w \leq_{i,s} v$ is to be read as: "at state $s$, agent $i$ considers $w$ at least as plausible as $v$". We take this relation to be a well-founded pre-order.[1] Also note that this relation is not only dependent on agents, but also on states: it is possible for agent $i$ to have different plausibility orderings at different states (from Section 4 onwards, more constraints will be placed on this state-dependency).

Various epistemic and doxastic notions can be interpreted on epistemic plausibility models. The three most important ones are: (i) $K_i \varphi$ (*i knows that* $\varphi$), (ii) $B_i^\alpha \varphi$ (*i believes that* $\varphi$, *conditional on* $\alpha$), and (iii) $B_i^+ \varphi$ (*i safely believes that* $\varphi$). 'Normal' belief can be defined in terms of conditional belief, by putting $B_i \varphi := B_i^\top \varphi$.

We abbreviate $[w]_{\sim_i} := \{v \in W \mid w \sim_i v\}$ (the $\sim_i$-equivalence class of state $w \in W$). The semantics for the notions above can now be stated as follows:

**Definition 2.** Consider an epistemic plausibility model $\mathbb{M}$ and state $w$; then

---

[1] For any state $s \in W$ and set $X \subseteq W$, we define $\mathrm{Min}_{\leq_{i,s}}(X) := \{x \in X \mid \forall y \in X : y \leq_{i,s} x \Rightarrow x \leq_{i,s} y\}$ (the set of $\leq_{i,w}$-minimal elements of $X$). That $\leq_{i,s}$ is a well-founded pre-order means that it is reflexive and transitive, and that for any $X \subseteq W$ such that $X \neq \emptyset$, also $\mathrm{Min}_{\leq_{i,s}}(X) \neq \emptyset$.

- $\mathbb{M}, w \models K_i\varphi$ iff $\forall v \in [w]_{\sim_i} : \mathbb{M}, v \models \varphi$
- $\mathbb{M}, w \models B_i^\alpha\varphi$ iff $\forall v \in W : v \in \text{Min}_{\leq_{i,w}}(\llbracket\alpha\rrbracket^\mathbb{M} \cap [w]_{\sim_i}) \Rightarrow \mathbb{M}, v \models \varphi$
- $\mathbb{M}, w \models B_i^+\varphi$ iff $\forall v \in [w]_{\sim_i} : v \leq_{i,w} w \Rightarrow \mathbb{M}, v \models \varphi$

We now turn to the dynamics. In this paper, we will focus on two specific dynamic phenomena: *public announcement* (hard information) and *radical upgrade* (soft information). Public announcement of a formula $\varphi$ in an epistemic plausibility model $\mathbb{M}$ simply removes all $\neg\varphi$-states from the model. Radical upgrade with $\varphi$, on the other hand, makes all $\varphi$-states more plausible than all $\neg\varphi$-states, and leaves everything within these two zones untouched. Formally, this looks as follows:

**Definition 3.** Consider an epistemic plausibility model $\mathbb{M} = \langle W, \{\sim_i\}_{i \in G}, \{\leq_{i,w}\}_{i \in G}^{w \in W}, V\rangle$ and a formula $\varphi$. We now define the following epistemic plausibility models:

- $\mathbb{M}!\varphi = \langle W^{!\varphi}, \{\sim_i^{!\varphi}\}_{i \in G}, \{\leq_{i,w}^{!\varphi}\}_{i \in G}^{w \in W^{!\varphi}}, V^{!\varphi}\rangle$, where
    - $W^{!\varphi} = \llbracket\varphi\rrbracket^\mathbb{M}$
    - $\sim_i^{!\varphi} := \sim_i \cap (\llbracket\varphi\rrbracket^\mathbb{M} \times \llbracket\varphi\rrbracket^\mathbb{M})$ for any $i \in G$
    - $\leq_{i,w}^{!\varphi} := \leq_{i,w} \cap (\llbracket\varphi\rrbracket^\mathbb{M} \times \llbracket\varphi\rrbracket^\mathbb{M})$ for any $i \in G$ and $w \in W^{!\varphi}$
    - $V^{!\varphi}(p) := V(p) \cap \llbracket\varphi\rrbracket^\mathbb{M}$ for any $p \in Prop$

- $\mathbb{M} \Uparrow \varphi = \langle W^{\Uparrow\varphi}, \{\sim_i^{\Uparrow\varphi}\}_{i \in G}, \{\leq_{i,w}^{\Uparrow\varphi}\}_{i \in G}^{w \in W^{\Uparrow\varphi}}, V^{\Uparrow\varphi}\rangle$, where
    - $W^{\Uparrow\varphi} := W$
    - $\sim_i^{\Uparrow\varphi} := \sim_i$ for any $i \in G$
    - $\leq_{i,w}^{\Uparrow\varphi} := \big(\leq_{i,w} \cap (\llbracket\varphi\rrbracket^\mathbb{M} \times \llbracket\varphi\rrbracket^\mathbb{M})\big) \cup \big(\leq_{i,w} \cap (\llbracket\neg\varphi\rrbracket^\mathbb{M} \times \llbracket\neg\varphi\rrbracket^\mathbb{M})\big) \cup \big(\llbracket\varphi\rrbracket^\mathbb{M} \times \llbracket\neg\varphi\rrbracket^\mathbb{M}\big)$ for any $i \in G$ and $w \in W^{\Uparrow\varphi}$
    - $V^{\Uparrow\varphi}(p) := V(p)$ for any $p \in Prop$

In order to be able to talk about these new models in the object language, we add operators $[!\varphi]$ and $[\Uparrow \varphi]$. Hence, the full language $\mathcal{L}(K, B^c, B^+, !, \Uparrow)$ has the following Backus-Naur Form (BNF):[2]

$$\begin{array}{rcl}\varphi &::=& p \mid \neg\varphi \mid \varphi \wedge \varphi \mid K_i\varphi \mid B_i^\varphi\varphi \mid B_i^+\varphi \mid [A]\varphi \\ A &::=& !\varphi \mid \Uparrow \varphi\end{array}$$

We now link up the models and the language by defining the semantics for the two dynamic modalities. Note that since public announcement is assumed to be *truthful*, it works with a precondition; this is not the case for radical upgrade.

**Definition 4.** Consider an epistemic plausibility model $\mathbb{M}$ and state $w$; then

- $\mathbb{M}, w \models [!\varphi]\psi$ iff (if $\mathbb{M}, w \models \varphi$ then $\mathbb{M}!\varphi, w \models \psi$)

---

[2] Of course, one can also study more restricted languages. BNF's for such restricted languages can easily be obtained from the BNF for the full language.

- $\mathbb{M}, w \models [\Uparrow \varphi]\psi$ iff $\mathbb{M} \Uparrow \varphi, w \models \psi$

Finally, dynamic epistemic/doxastic logics are constructed using the well-known *modular approach*: (i) one starts by taking (an axiomatization of) some static base logic (in a sufficiently rich language, so that step (iii) can be done successfully[3]), (ii) then one adds dynamic operators to this logic and (iii) finally, one provides a sound set of reduction axioms, which allow each formula in the dynamic language to be rewritten as an equivalent formula in the static language. Because of this final step, completeness of the dynamified logic is reduced to completeness of the static base logic. It also shows that the dynamic language $\mathcal{L}(K, B^c, B^+, !, \Uparrow)$ is equally expressive as the static language $\mathcal{L}(K, B^c, B^+)$.

We illustrate this methodology by providing the most important reduction axioms for public announcement and radical upgrade, viz. those in which the epistemic/doxastic operators are being rewritten:

**Fact 5.** The following are all sound with respect to epistemic plausibility models:

$$
\begin{array}{lcl}
[!\varphi]K_i\psi & \leftrightarrow & (\varphi \to K_i[!\varphi]\psi) \\
[!\varphi]B_i^\alpha\psi & \leftrightarrow & (\varphi \to B_i^{\varphi \wedge [!\varphi]\alpha}[!\varphi]\psi) \\
[!\varphi]B_i^+\psi & \leftrightarrow & (\varphi \to B_i^+[!\varphi]\psi) \\
[\Uparrow \varphi]K_i\psi & \leftrightarrow & K_i[\Uparrow \varphi]\psi \\
[\Uparrow \varphi]B_i^\alpha\psi & \leftrightarrow & \big(\hat{K}_i(\varphi \wedge [\Uparrow \varphi]\alpha) \wedge B_i^{\varphi \wedge [\Uparrow \varphi]\alpha}[\Uparrow \varphi]\psi\big) \vee \\
 & & \big(\neg \hat{K}_i(\varphi \wedge [\Uparrow \varphi]\alpha) \wedge B_i^{[\Uparrow \varphi]\alpha}[\Uparrow \varphi]\psi\big) \\
[\Uparrow \varphi]B_i^+\psi & \leftrightarrow & \big(\varphi \to B_i^+(\varphi \to [\Uparrow \varphi]\psi)\big) \wedge \\
 & & \big(\neg \varphi \to (B_i^+(\neg \varphi \to [\Uparrow \varphi]\psi) \wedge K_i(\varphi \to [\Uparrow \varphi]\psi))\big)
\end{array}
$$

## 3 Bisimulation for epistemic plausibility models

We now start our investigation of the model theory of epistemic plausibility models. The focus will be on the notion of *bisimulation*, which is also central in the model theory of Kripke models. Since we want to explore bisimulation for various languages, we make it into a parametrized notion, so that each language has its own notion of bisimulation, which 'does what it needs to do, and nothing more'.

Below are the definitions of $K$-bisimulation, $B^+$-bisimulation and $B^c$-bisimulation. Since $K_i$ is just the universal modality for $\sim_i$, the notion of $K$-bisimulation is that of regular bisimulation from modal logic. The notion of $B^+$-bisimulation is a straightforward generalization. The notion of $B^c$-bisimulation, however, is much more intricate, since it involves universally quantifying over all formulas of the language $\mathcal{L}(B^+)$. We will return to this issue in later sections.

**Definition 6.** Given epistemic plausiblity models $\mathbb{M} = \langle W, \{\sim_i\}_{i \in G}, \{\leq_{i,w}\}_{i \in G}^{w \in W}, V\rangle$ and $\mathbb{M}' = \langle W', \{\sim_i'\}_{i \in G}, \{\leq_{i,w'}'\}_{i \in G}^{w' \in W'}, V'\rangle$ ; a relation $Z \subseteq W \times W'$ is a $K$-*bisimulation* iff

- if $(w, w') \in Z$, then for all atoms $p$: $w \in V(p) \Leftrightarrow w' \in V'(p)$

- if $(w, w') \in Z$ and $w \sim_i v$, then there is a $v' \in W'$ such that $(v, v') \in Z$ and $w' \sim_i' v'$

- if $(w, w') \in Z$ and $w' \sim_i' v'$, then there is a $v \in W$ such that $(v, v') \in Z$ and $w \sim_i v$

---

[3] For example, if the language contains radical upgrade and safe belief operators, then it should also contain the knowledge operator, so that the reduction axiom is expressible; cf. Fact 5.

**Definition 7.** Given epistemic plausiblity models $\mathbb{M} = \langle W, \{\sim_i\}_{i \in G}, \{\leq_{i,w}\}_{i \in G}^{w \in W}, V\rangle$ and $\mathbb{M}' = \langle W', \{\sim'_i\}_{i \in G}, \{\leq'_{i,w'}\}_{i \in G}^{w' \in W'}, V'\rangle$ ; a relation $Z \subseteq W \times W'$ is a $B^+$-*bisimulation* iff

- if $(w, w') \in Z$, then for all atoms $p$: $w \in V(p) \Leftrightarrow w' \in V'(p)$

- if $(w, w') \in Z$ and $w \sim_i v$ and $v \leq_{i,w} w$, then there is a $v' \in W'$ such that $(v, v') \in Z$ and $w' \sim'_i v'$ and $v' \leq'_{i,w'} w'$

- if $(w, w') \in Z$ and $w' \sim'_i v'$ and $v' \leq'_{i,w'} w'$, then there is a $v \in W$ such that $(v, v') \in Z$ and $w \sim_i v$ and $v \leq_{i,w} w$

**Definition 8.** Given epistemic plausiblity models $\mathbb{M} = \langle W, \{\sim_i\}_{i \in G}, \{\leq_{i,w}\}_{i \in G}^{w \in W}, V\rangle$ and $\mathbb{M}' = \langle W', \{\sim'_i\}_{i \in G}, \{\leq'_{i,w'}\}_{i \in G}^{w' \in W'}, V'\rangle$ ; a relation $Z \subseteq W \times W'$ is a $B^c$-*bisimulation* iff

- if $(w, w') \in Z$, then for all atoms $p$: $w \in V(p) \Leftrightarrow w' \in V'(p)$

- for all formulas $\alpha \in \mathcal{L}(B^c)$: if $(w, w') \in Z$ and $v \in \text{Min}_{\leq_{i,w}}([\![\alpha]\!]^{\mathbb{M}} \cap [w]_{\sim_i})$, then there is a $v' \in W'$ such that $(v, v') \in Z$ and $v' \in \text{Min}_{\leq'_{i,w'}}([\![\alpha]\!]^{\mathbb{M}'} \cap [w']_{\sim'_i})$

- for all formulas $\alpha \in \mathcal{L}(B^c)$: if $(w, w') \in Z$ and $v' \in \text{Min}_{\leq'_{i,w'}}([\![\alpha]\!]^{\mathbb{M}'} \cap [w']_{\sim'_i})$, then there is a $v \in W$ such that $(v, v') \in Z$ and $v \in \text{Min}_{\leq_{i,w}}([\![\alpha]\!]^{\mathbb{M}} \cap [w]_{\sim_i})$

The following theorem shows that these are the 'right' notions, since they allow us to establish a characteristic feature of bisimulation: bisimilarity implies modal equivalence.

**Theorem 9.** Consider two epistemic plausiblity models $\mathbb{M} = \langle W, \{\sim_i\}_{i \in G}, \{\leq_{i,w}\}_{i \in G}^{w \in W}, V\rangle$ and $\mathbb{M}' = \langle W', \{\sim'_i\}_{i \in G}, \{\leq'_{i,w'}\}_{i \in G}^{w' \in W'}, V'\rangle$, and a relation $Z \subseteq W \times W'$.

1. If $Z$ is a $K$-bisimulation, then for all $\varphi \in \mathcal{L}(K)$ and for all $(w, w') \in Z$, it holds that $\mathbb{M}, w \models \varphi \Leftrightarrow \mathbb{M}', w' \models \varphi$.

2. If $Z$ is a $B^+$-bisimulation, then for all $\varphi \in \mathcal{L}(B^+)$ and for all $(w, w') \in Z$, it holds that $\mathbb{M}, w \models \varphi \Leftrightarrow \mathbb{M}', w' \models \varphi$.

3. If $Z$ is a $B^c$-bisimulation, then for all $\varphi \in \mathcal{L}(B^c)$ and for all $(w, w') \in Z$, it holds that $\mathbb{M}, w \models \varphi \Leftrightarrow \mathbb{M}', w' \models \varphi$.

*Proof.* Each of these three statements is easily proved by induction on the complexity of $\varphi$. To illustrate this, we treat the cases for the epistemic/doxastic modality.

1. Suppose that $(w, w') \in Z$ and $\mathbb{M}, w \models K_i\varphi$; we show that also $\mathbb{M}', w' \models K_i\varphi$. Consider an arbitrary $v' \in W'$ and suppose that $w' \sim'_i v'$. By Definition 6, there exists a $v \in W$ such that $(v, v') \in Z$ and $w \sim_i v$. Since $\mathbb{M}, w \models K_i\varphi$ and $w \sim_i v$, we get $\mathbb{M}, v \models \varphi$. Since $(v, v') \in Z$, we get by the induction hypothesis that also $\mathbb{M}', v' \models \varphi$. The other direction is completely analogous.

2. Suppose that $(w, w') \in Z$ and $\mathbb{M}, w \models B_i^+\varphi$; we show that also $\mathbb{M}', w' \models B_i^+\varphi$. Consider an arbitrary $v' \in W'$ and suppose that $w' \sim'_i v'$ and $v' \leq'_{i,w'} w'$. By Definition 7, there exists a $v \in W$ such that $(v, v') \in Z$ and $w \sim_i v$ and $v \leq_{i,w} w$. Since $\mathbb{M}, w \models B_i^+\varphi$ and $w \sim_i v$ and $v \leq_{i,w} w$, we get $\mathbb{M}, v \models \varphi$. Since $(v, v') \in Z$, we get by the induction hypothesis that also $\mathbb{M}', v' \models \varphi$. The other direction is completely analogous.

3. Suppose that $(w,w') \in Z$ and $\mathbb{M}, w \models B_i^\alpha \varphi$; we show that also $\mathbb{M}', w' \models B_i^\alpha \varphi$. Consider an arbitrary $v' \in W'$ and suppose that $v' \in \text{Min}_{\leq'_{i,w'}}(\llbracket \alpha \rrbracket^{\mathbb{M}'} \cap [w']_{\sim'_i})$. Since $B_i^\alpha \varphi \in \mathcal{L}(B^c)$, also $\alpha \in \mathcal{L}(B^c)$; hence, by Definition 8, there exists a $v \in W$ such that $(v, v') \in Z$ and $v \in \text{Min}_{\leq_{i,w}}(\llbracket \alpha \rrbracket^{\mathbb{M}} \cap [w]_{\sim_i})$. Since $\mathbb{M}, w \models B_i^\alpha \varphi$ and $v \in \text{Min}_{\leq_{i,w}}(\llbracket \alpha \rrbracket^{\mathbb{M}} \cap [w]_{\sim_i})$, we get $\mathbb{M}, v \models \varphi$. Since $(v, v') \in Z$, we get by the induction hypothesis that also $\mathbb{M}', v' \models \varphi$. The other direction is completely analogous. □

Using these separate notions of bisimulations, we can now introduce bisimulations for languages which have more than just one of the operators $K/B^+/B^c$ in a modular way (although conditional belief complicates matters a little bit). Obviously, these combined notions lead to results analogous to Theorem 9; we state just two of these (without proof) as Theorem 11, for future reference.

**Definition 10.** Consider epistemic plausiblity models $\mathbb{M} = \langle W, \{\sim_i\}_{i \in G}, \{\leq_{i,w}\}_{i \in G}^{w \in W}, V \rangle$ and $\mathbb{M}' = \langle W', \{\sim'_i\}_{i \in G}, \{\leq'_{i,w'}\}_{i \in G}^{w' \in W'}, V' \rangle$ and a relation $Z \subseteq W \times W'$.

- $Z$ is a $\{K, B^+\}$-bisimulation iff $Z$ is a $K$-bisimulation and a $B^+$-bisimulation

- $Z$ is a $\{K, B^c\}$-bisimulation iff $Z$ is a $K$-bisimulation and a $B^c$-bisimulation, with the universal quantifiers in Definition 8 ranging over $\mathcal{L}(K, B^c)$ (instead of just over $\mathcal{L}(B^c)$)

- $Z$ is a $\{K, B^+, B^c\}$-bisimulation iff $Z$ is a $K$-bisimulation, a $B^+$-bisimulation, and a $B^c$-bisimulation, with the universal quantifiers in Definition 8 ranging over $\mathcal{L}(K, B^+, B^c)$ (instead of just over $\mathcal{L}(B^c)$)

**Theorem 11.** Consider two epistemic plausiblity models $\mathbb{M} = \langle W, \{\sim_i\}_{i \in G}, \{\leq_{i,w}\}_{i \in G}^{w \in W}, V \rangle$ and $\mathbb{M}' = \langle W', \{\sim'_i\}_{i \in G}, \{\leq'_{i,w'}\}_{i \in G}^{w' \in W'}, V' \rangle$, and a relation $Z \subseteq W \times W'$.

1. If $Z$ is a $\{K, B^c\}$-bisimulation, then for all $\varphi \in \mathcal{L}(K, B^c)$ and for all $(w, w') \in Z$, it holds that $\mathbb{M}, w \models \varphi \Leftrightarrow \mathbb{M}', w' \models \varphi$.

2. If $Z$ is a $\{K, B^+\}$-bisimulation, then for all $\varphi \in \mathcal{L}(K, B^+)$ and for all $(w, w') \in Z$, it holds that $\mathbb{M}, w \models \varphi \Leftrightarrow \mathbb{M}', w' \models \varphi$.

One can also wonder about the converse direction of theorems such as Theorem 11: if $\mathbb{M}, w \models \varphi \Leftrightarrow \mathbb{M}', w' \models \varphi$ for all $\varphi \in \mathcal{L}(K, B^c)$, then is there always a $\{K, B^c\}$-bisimulation $Z \subseteq W \times W'$ such that $(w, w') \in Z$? One of the main results from the model theory of basic modal logic, viz. the Hennesy-Milner theorem (cf. [2], Theorem 2.24) says that this question can be answered positively, at least when the models are assumed to be image-finite. This theorem can easily be generalized to epistemic plausibility models:

**Definition 12.** Consider an epistemic plausibility model $\mathbb{M} = \langle W, \{\sim_i\}_{i \in G}, \{\leq_{i,w}\}_{i \in G}^{w \in W}, V \rangle$. We say that $\mathbb{M}$ is *image-finite* if for all $i \in G$ and all $w \in W$, the set $[w]_{\sim_i}$ is finite.

**Theorem 13.** Consider two image-finite models $\mathbb{M} = \langle W, \{\sim_i\}_{i \in G}, \{\leq_{i,w}\}_{i \in G}^{w \in W}, V \rangle$ and $\mathbb{M}' = \langle W', \{\sim'_i\}_{i \in G}, \{\leq'_{i,w'}\}_{i \in G}^{w' \in W'}, V' \rangle$. Then for all states $w \in W$ and $w' \in W'$, if $\mathbb{M}, w \models \varphi \Leftrightarrow \mathbb{M}', w' \models \varphi$ for all $\varphi \in \mathcal{L}(K, B^c)$ (we will write $w \equiv_{\{K, B^c\}} w'$), then $w$ and $w'$ are $\{K, B^c\}$-bisimilar.

*Proof.* We use the trick of the Hennessy-Milner theorem for basic modal logic, viz. we show that $\equiv_{\{K,B^c\}}$ is itself a $\{K, B^c\}$-bisimulation. If $w \equiv_{\{K,B^c\}} w'$, then the atoms clause is trivially fulfilled. We now focus on the zig-clauses of Definitions 6 and 8 (extended to $\mathcal{L}(K, B^c)$ instead of just $\mathcal{L}(B^c)$, cf. supra); the zag-clauses are treated completely analogously.

*Zig-clause for $K$-bisimulation.* Suppose that $w \equiv_{\{K,B^c\}} w'$ and $w \sim_i v$; we will show that there is a $v' \in W'$ such that $v \equiv_{\{K,B^c\}} v'$ and $w' \sim'_i v'$. For a reductio, suppose that this is not the case; so for all $v' \in [w']_{\sim'_i}$ we have $v \not\equiv_{\{K,B^c\}} v'$ (*). If $[w']_{\sim'_i} = \emptyset$, then $\mathbb{M}', w' \models K_i \bot$, so since $w \equiv_{\{K,B^c\}} w'$, we get $\mathbb{M}, w \models K_i \bot$, which contradicts $v \in [w]_{\sim_i}$. Hence $[w']_{\sim'_i} \neq \emptyset$. Since $\mathbb{M}'$ is image-finite, $[w']_{\sim'_i}$ is finite, say $[w']_{\sim'_i} = \{v'_1, \ldots, v'_n\}$. We can now rephrase (*) as follows: for all $k \in \{1, \ldots, n\}$, there is a $\varphi_k \in \mathcal{L}(K, B^c)$ such that $\mathbb{M}, v \models \varphi_k$ and $\mathbb{M}', v'_k \not\models \varphi_k$. It now easily follows that $\mathbb{M}, w \models \neg K_i \neg(\varphi_1 \wedge \cdots \wedge \varphi_n)$, and yet $\mathbb{M}', w' \not\models \neg K_i \neg(\varphi_1 \wedge \cdots \wedge \varphi_n)$. Since $\varphi_1, \ldots, \varphi_n \in \mathcal{L}(K, B^c)$, also $\neg K_i \neg(\varphi_1 \wedge \cdots \wedge \varphi_n) \in \mathcal{L}(K, B^c)$. This contradicts $w \equiv_{\{K,B^c\}} w'$.

*Zig-clause for $B^c$-bisimulation.* Consider an arbitrary formula $\alpha \in \mathcal{L}(K, B^c)$ and suppose that $w \equiv_{\{K,B^c\}} w'$ and $v \in \text{Min}_{\leq_{i,w}}(\llbracket \alpha \rrbracket^{\mathbb{M}} \cap [w]_{\sim_i})$. We will now show that there is a $v' \in W'$ such that $v \equiv_{\{K,B^c\}} v'$ and $v' \in \text{Min}_{\leq'_{i,w'}}(\llbracket \alpha \rrbracket^{\mathbb{M}'} \cap [w']_{\sim'_i})$. For a reductio, suppose that this is not the case; so for all $v' \in \text{Min}_{\leq'_{i,w'}}(\llbracket \alpha \rrbracket^{\mathbb{M}'} \cap [w']_{\sim'_i})$ we have $v \not\equiv_{\{K,B^c\}} v'$ (**). If $\text{Min}_{\leq'_{i,w'}}(\llbracket \alpha \rrbracket^{\mathbb{M}'} \cap [w']_{\sim'_i}) = \emptyset$, then $\mathbb{M}', w' \models B_i^\alpha \bot$, so since $w \equiv_{\{K,B^c\}} w'$, we get $\mathbb{M}, w \models B_i^\alpha \bot$, which contradicts $v \in \text{Min}_{\leq_{i,w}}(\llbracket \alpha \rrbracket^{\mathbb{M}} \cap [w]_{\sim_i})$. Hence $\text{Min}_{\leq'_{i,w'}}(\llbracket \alpha \rrbracket^{\mathbb{M}'} \cap [w']_{\sim'_i}) \neq \emptyset$. Since $\mathbb{M}'$ is image-finite, $[w']_{\sim'_i}$ is finite, so $\llbracket \alpha \rrbracket^{\mathbb{M}'} \cap [w']_{\sim'_i}$ is also finite, and so $\text{Min}_{\leq'_{i,w'}}(\llbracket \alpha \rrbracket^{\mathbb{M}'} \cap [w']_{\sim'_i})$ is also finite — say $\text{Min}_{\leq'_{i,w'}}(\llbracket \alpha \rrbracket^{\mathbb{M}'} \cap [w']_{\sim'_i}) = \{v'_1, \ldots, v'_m\}$. We can now rephrase (**) as follows: for all $k \in \{1, \ldots, m\}$, there is a $\varphi_k \in \mathcal{L}(K, B^c)$ such that $\mathbb{M}, v \models \varphi_k$ and $\mathbb{M}', v'_k \not\models \varphi_k$. It now easily follows that $\mathbb{M}, w \models \neg B_i^\alpha \neg(\varphi_1 \wedge \cdots \wedge \varphi_m)$, and yet $\mathbb{M}', w' \not\models \neg B_i^\alpha \neg(\varphi_1 \wedge \cdots \wedge \varphi_m)$. Since $\alpha, \varphi_1, \ldots, \varphi_m \in \mathcal{L}(K, B^c)$, also $\neg B_i^\alpha \neg(\varphi_1 \wedge \cdots \wedge \varphi_m) \in \mathcal{L}(K, B^c)$. This contradicts $w \equiv_{\{K,B^c\}} w'$. □

One of the main uses of bisimulation is studying (un)definability results. We will now illustrate this by proving two undefinability theorems. Both of these theorems can be seen as tying up some loose ends, in the sense that the results were expected, but not yet explicitly proved in the existing literature.

**Theorem 14.** Conditional belief cannot be defined in terms of knowledge and safe belief.[4]

*Proof.* For a reductio, suppose conditional belief *is* definable in terms of knowledge and safe belief. Consider the formula $B^p q$. Since conditional belief is definable in terms of $K$ and $B^+$, there is a formula $\varphi \in \mathcal{L}(K, B^+)$ such that $\mathbb{M}, w \models B^p q \leftrightarrow \varphi$ for all epistemic plausibility models $\mathbb{M}$ and states $w$.

Now consider the models $\mathbb{M}$ and $\mathbb{M}'$ pictured below (we focus on one agent $i$, and drop agent subscripts for the sake of readability). It is easy to check that the dotted line $Z$ is a $\{K, B^+\}$-bisimulation. Since $\varphi \in \mathcal{L}(K, B^+)$, it follows by Theorem 9 that $\mathbb{M}, w \models \varphi$ iff $\mathbb{M}', w' \models \varphi$, and hence also $\mathbb{M}, w \models B^p q$ iff $\mathbb{M}', w' \models B^p q$.

---

[4]This theorem does not contradict Theorem 27, since that definability theorem holds for a *restricted* class of epistemic plausibility models, whereas this undefinability theorem holds for the *entire* class of epistemic plausibility models; also note that the 'counterexample models' used to prove this theorem do not belong to the restricted class of Theorem 27.

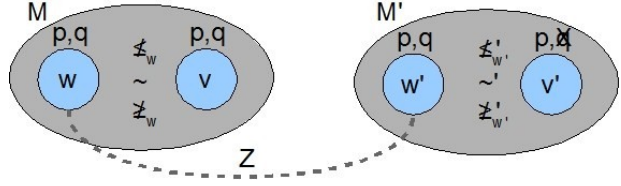

Since $w \not\leq_w v$ and $v \not\leq_w w$, we have $w, v \in \text{Min}_{\leq_w}\{w, v\}$, and hence $\text{Min}_{\leq_w}(\llbracket p \rrbracket^M \cap [w]_\sim) = \text{Min}_{\leq_w}\{w, v\} = \{w, v\}$. Analogously $\text{Min}_{\leq_{w'}'}(\llbracket p \rrbracket^{M'} \cap [w']_{\sim'}) = \{w', v'\}$. Since $\mathbb{M}, w \models q$ and $\mathbb{M}, v \models q$, we get $\mathbb{M}, w \models B^p q$, but since $\mathbb{M}', v' \not\models q$, we get $\mathbb{M}', w' \not\models B^p q$; contradiction. $\square$

**Theorem 15.** *Safe belief cannot be defined in terms of knowledge and conditional belief.*

*Proof.* For a reductio, suppose safe belief *is* definable in terms of knowledge and conditional belief. Consider the formula $B^+ p$. Since safe belief is definable in terms of knowledge and conditional belief, there is a formula $\varphi \in \mathcal{L}(K, B^c)$ such that $\mathbb{M}, w \models B^+ p \leftrightarrow \varphi$ for all epistemic plausibility models $\mathbb{M}$ and states $w$.

Now consider the models $\mathbb{M}$ and $\mathbb{M}'$ pictured below (we focus on one agent $i$, and drop agent subscripts for the sake of readability). It is easy to check that the dotted lines $Z$ form a $K$-bisimulation. We claim that they also form a $B^c$-bisimulation (this claim is proved later). Since $\varphi \in \mathcal{L}(K, B^c)$, it follows by the first part of Theorem 11 that $\mathbb{M}, w \models \varphi$ iff $\mathbb{M}', w' \models \varphi$, and hence also $\mathbb{M}, w \models B^+ p$ iff $\mathbb{M}', w' \models B^+ p$. However, using Definition 2, one easily checks that $\mathbb{M}, w \models B^+ p$, while $\mathbb{M}', w' \not\models B^+ p$; contradiction.

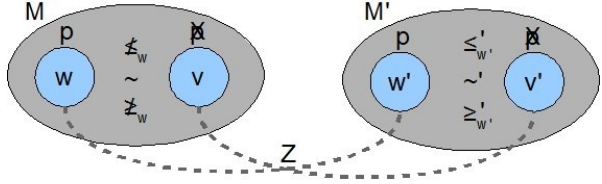

We now prove the claim that $Z$ is a bisimulation. Note that for any $x \in \{w, v\}$ and $X \subseteq \{w, v\}$, it holds that $\text{Min}_{\leq_x}(X) = X$ and $[x]_\sim = \{w, v\}$.[5] Hence we get that $y \in \text{Min}_{\leq_x}(\llbracket \alpha \rrbracket^{\mathbb{M}} \cap [x]_\sim)$ iff $y \in \llbracket \alpha \rrbracket^{\mathbb{M}} \cap [x]_\sim$ iff $\mathbb{M}, y \models \alpha$. Analogously, we show that $y' \in \text{Min}_{\leq_{x'}'}(\llbracket \alpha \rrbracket^{\mathbb{M}'} \cap [x']_{\sim'})$ iff $\mathbb{M}', y' \models \alpha$. We already know that $w$ and $w'$, and $v$ and $v'$ agree on all atoms. Hence, reconsidering Definition 8, the following remains to be shown:

$\forall \alpha \in \mathcal{L}(K, B^c)$: if $(x, x') \in Z$ and $\mathbb{M}, y \models \alpha$, then $\exists y' \in W' : (y, y') \in Z$ and $\mathbb{M}', y' \models \alpha$
$\forall \alpha \in \mathcal{L}(K, B^c)$: if $(x, x') \in Z$ and $\mathbb{M}', y' \models \alpha$, then $\exists y \in W : (y, y') \in Z$ and $\mathbb{M}, y \models \alpha$

Note that the condition $(x, x') \in Z$ is vacuous. Furthermore, since $Z = \{(w, w'), (v, v')\}$, this can be rewritten as the following

**Claim.** $\forall \alpha \in \mathcal{L}(K, B^c) : \mathbb{M}, w \models \alpha \Leftrightarrow \mathbb{M}', w' \models \alpha$ and $\mathbb{M}, v \models \alpha \Leftrightarrow \mathbb{M}', v' \models \alpha$

This claim is easily proved by induction on the complexity of $\alpha$. We treat the cases for knowledge and conditional belief. If $\alpha = K\varphi$, then we have:

---
[5] Although this is not explicitly represented in the picture, we are assuming that $\leq_v = \{(w, w), (v, v)\}$ and $\leq_{v'}' = \{(w', w'), (v', v')\}$.

$$\begin{aligned}
\mathbb{M}, w \models K\varphi &\Leftrightarrow \forall x \in [w]_\sim : \mathbb{M}, x \models \varphi \\
&\Leftrightarrow \mathbb{M}, w \models \varphi \text{ and } \mathbb{M}, v \models \varphi \\
&\Leftrightarrow \mathbb{M}', w' \models \varphi \text{ and } \mathbb{M}', v' \models \varphi \quad \text{(IH)} \\
&\Leftrightarrow \forall x' \in [w']_{\sim'} : \mathbb{M}', x' \models \varphi \\
&\Leftrightarrow \mathbb{M}', w' \models K\varphi
\end{aligned}$$

and similarly $\mathbb{M}, v \models K\varphi \Leftrightarrow \mathbb{M}', v' \models K\varphi$. If $\alpha = B^\psi \varphi$, then we have:

$$\begin{aligned}
\mathbb{M}, w \models B^\psi\varphi &\Leftrightarrow \forall x \in \mathrm{Min}_{\leq_w}(\llbracket\psi\rrbracket^\mathbb{M} \cap [w]_\sim) : \mathbb{M}, x \models \varphi \\
&\Leftrightarrow \forall x \in \llbracket\psi\rrbracket^\mathbb{M} : \mathbb{M}, x \models \varphi & \text{(cf. supra)} \\
&\Leftrightarrow \forall x \in W : \mathbb{M}, x \models \psi \Rightarrow \mathbb{M}, x \models \varphi \\
&\Leftrightarrow \mathbb{M}, w \models \psi \Rightarrow \mathbb{M}, w \models \varphi \text{ and } \mathbb{M}, v \models \psi \Rightarrow \mathbb{M}, v \models \varphi \\
&\Leftrightarrow \mathbb{M}', w' \models \psi \Rightarrow \mathbb{M}', w' \models \varphi \text{ and } \mathbb{M}', v' \models \psi \Rightarrow \mathbb{M}', v' \models \varphi & \text{(IH)} \\
&\Leftrightarrow \forall x' \in W' : \mathbb{M}', x' \models \psi \Rightarrow \mathbb{M}', x' \models \varphi \\
&\Leftrightarrow \forall x' \in \llbracket\psi\rrbracket^{\mathbb{M}'} : \mathbb{M}', x' \models \varphi \\
&\Leftrightarrow \forall x' \in \mathrm{Min}_{\leq'_{w'}}(\llbracket\psi\rrbracket^{\mathbb{M}'} \cap [w']_{\sim'}) : \mathbb{M}', x' \models \varphi & \text{(cf. supra)} \\
&\Leftrightarrow \mathbb{M}, w \models B^\psi\varphi
\end{aligned}$$

and similarly $\mathbb{M}, v \models B^\psi\varphi \Leftrightarrow \mathbb{M}', v' \models B^\psi\varphi$ (recall Footnote 5). This finishes the proof of the claim, and thus also of Theorem 15. □

## 4 Structural bisimulations

We already noted in the previous section that the notion of $B^c$-bisimulation introduced in Definition 8 is much more intricate than the other notions. We will now argue that this definition is unsatisfactory for both theoretical and practical reasons.

On the *theoretical* level, since Definition 8 involves universal quantification over $\mathcal{L}(B^c)$, it is not strictly structural. Rather than stating conditions on $\sim_i$ and $\leq_{i,w}$ (as is done in Definitions 6 and 7 of bisimulations for knowledge and safe belief), it essentially involves truth sets of (arbitrary) formulas. A related issue is that this definition of bisimilarity for *models* cannot be turned into a definition of bisimilarity for *frames* by simply dropping the 'atoms' clause (as can be done with Definitions 6 and 7): it depends on truth sets of formulas ($\llbracket\alpha\rrbracket^\mathbb{M}$ and $\llbracket\alpha\rrbracket^{\mathbb{M}'}$), and thus also on the concrete valuations of the models $\mathbb{M}$ and $\mathbb{M}'$.

*Practically* speaking, Definition 8 makes it often very difficult to prove that two given epistemic plausibility models are actually $B^c$-bisimilar. In the appendix of [4], induction on the complexity of $\alpha$ (with a cleverly strengthened induction hypothesis) is used to establish that the zig- and zag-conditions of Definition 8 hold for all formulas $\alpha$. However, this approach is geared towards proving one particular $B^c$-bisimilarity result (about two artificially crafted models), and cannot easily be generalized to the general case (proving $B^c$-bisimilarity of arbitrary models). Similar remarks apply to our proof of Theorem 15. Furthermore, recall that one of the main goals of introducing bisimulations is that they allow us to prove equivalence results. For example, in Theorem 15, we want to show that the two pictured models are $\{K, B^c\}$-bisimilar, and then (using the first part of Theorem 11) conclude that they are $\mathcal{L}(K, B^c)$-equivalent. However, note that while establishing the $B^c$-bisimilarity, we ended up proving a separate claim, which just *is* the original $\mathcal{L}(K, B^c)$-equivalence result we were looking for. This seems to be some kind of practical 'circularity' (we want bisimilarity to get equivalence — but to get bisimilarity, we already need equivalence), which renders the current notion of $B^c$-bisimulation practically useless.

We will now propose two different solutions to this problem, and explore and compare their advantages and disadvantages. Both solutions involve reducing conditional belief to other modalities which have more standard notions of bisimulation. The first approach involves both extending the language and putting some mild constraints on the epistemic plausibility models. The second approach puts more heavy constraints on the models, but does not need to extend the language. Both solutions have in common that we end up only needing fully structural notions of bisimulation, without any universal quantification over formulas.

### 4.1 Adding a new modality

The first approach[6] combines language engineering and putting some mild constraints on the models. These constraints are captured by the following definition:

**Definition 16.** An epistemic plausibility model $\mathbb{M} = \langle W, \{\sim_i\}_{i \in G}, \{\leq_{i,w}\}_{i \in G}^{w \in W}, V \rangle$ is called *uniform* iff the plausibility relations are uniform within epistemic equivalence classes, i.e. iff for any $i \in G$ and $w, v \in W$: if $w \sim_i v$ then $\leq_{i,w} = \leq_{i,v}$.

This is a natural condition to impose on epistemic plausibility models: it leads to the (intuitively plausible) epistemic/doxastic introspection principle that agents know their (conditional) beliefs. Furthermore, uniformity is a *dynamically robust* notion, in the sense that if an epistemic plausibility model is uniform, then after it has undergone some dynamics, it is still uniform.

**Theorem 17.** If an epistemic plausibility model $\mathbb{M}$ is uniform, then $\mathbb{M} \models B_i^\alpha \varphi \to K_i B_i^\alpha \varphi$.

*Proof.* Consider an arbitrary state $w$ of $\mathbb{M}$ and suppose that $\mathbb{M}, w \models B_i^\alpha \varphi$. Consider an arbitrary $v \in [w]_{\sim_i}$; it now suffices to show that $\mathbb{M}, v \models B_i^\alpha \varphi$. Since $\mathbb{M}, w \models B_i^\alpha \varphi$, we have $\text{Min}_{\leq_{i,w}}(\llbracket \alpha \rrbracket^\mathbb{M} \cap [w]_{\sim_i}) \subseteq \llbracket \varphi \rrbracket^\mathbb{M}$ (†). Since $w \sim_i v$, we have $[w]_{\sim_i} = [v]_{\sim_i}$ (because $\sim_i$ is an equivalence relation) and $\leq_{i,w} = \leq_{i,v}$ (because $\mathbb{M}$ is uniform). Hence (†) becomes: $\text{Min}_{\leq_{i,v}}(\llbracket \alpha \rrbracket^\mathbb{M} \cap [v]_{\sim_i}) \subseteq \llbracket \varphi \rrbracket^\mathbb{M}$, ergo $\mathbb{M}, v \models B_i^\alpha \varphi$. □

**Theorem 18.** If an epistemic plausibility model $\mathbb{M}$ is uniform, then so are $\mathbb{M}!\varphi$ and $\mathbb{M} \Uparrow \varphi$.

*Proof.* Consider a uniform model $\mathbb{M}$. We treat the case of public announcement. Consider arbitrary states $w, v$ of $\mathbb{M}!\varphi$ and suppose that $w \sim_i^{!\varphi} v$. By Definition 3, we get that $w \sim_i v$. Since $\mathbb{M}$ is uniform, it follows that $\leq_{i,w} = \leq_{i,v}$. Hence also $\leq_{i,w}^{!\varphi} = \leq_{i,w} \cap (\llbracket \varphi \rrbracket^\mathbb{M} \times \llbracket \varphi \rrbracket^\mathbb{M}) = \leq_{i,v} \cap (\llbracket \varphi \rrbracket^\mathbb{M} \times \llbracket \varphi \rrbracket^\mathbb{M}) = \leq_{i,v}^{!\varphi}$. The case of radical upgrade is completely analogous. □

Uniform epistemic plausibility models will become very important later on. First, however, we need to set up some other things. For any agent $i \in G$ and state $w$ in a plausibility model, let us abbreviate $<_{i,w} := \leq_{i,w} - \geq_{i,w}$ and $\cong_{i,w} := \leq_{i,w} \cap \geq_{i,w}$ (so $x <_{i,w} y$ iff $x \leq_{i,w} y$ and not $y \leq_{i,w} x$; and $x \cong_{i,w} y$ iff $x \leq_{i,w} y$ and $y \leq_{i,w} x$). Note that since $\leq_{i,w}$ is a pre-order and thus not necessarily antisymmetric, it is possible that $x \cong_{i,w} y$ and yet $x \neq y$. One can easily verify the following fact, which expresses minimality in terms of the strict ordering $<$:

**Fact 19.** Consider an epistemic plausibility model $\mathbb{M} = \langle W, \{\sim_i\}_{i \in G}, \{\leq_{i,w}\}_{i \in G}^{w \in W}, V \rangle$, a set $X \subseteq W$ and a state $w \in W$. Then for any state $x \in W$, it holds that $x \in \text{Min}_{\leq_{i,w}}(X)$ iff $x \in X$ and there is no $y \in X$ such that $y <_{i,w} x$.

---

[6]This approach is based on a suggestion by Johan van Benthem and Davide Grossi.

We now extend our language with a modality $[>_i]$ to talk about this strict version of the plausibility order. As in Definition 2, the semantics for this modality is relativized to the epistemic equivalence classes:

**Definition 20.** Consider an epistemic plausibility model $\mathbb{M}$ and state $w$; then
$$\mathbb{M}, w \models [>_i]\varphi \text{ iff } \forall v \in [w]_{\sim_i} : v <_{i,w} w \Rightarrow \mathbb{M}, v \models \varphi$$

We now show that adding this new modality $[>_i]$ as a primitive operator is justified, in the sense that it cannot be defined in even the richest language of the previous section:

**Theorem 21.** The modality $[>_i]$ cannot be defined in $\mathcal{L}(K, B^c, B^+)$.

*Proof.* We focus on one agent $i$, and drop agent subscripts for the sake of readability. Consider the models pictured below (we assume that all proposition letters are made true everywhere: $V(p) = \{w, v\}, V'(p) = \{w', v'\}$ for all $p \in Prop$).

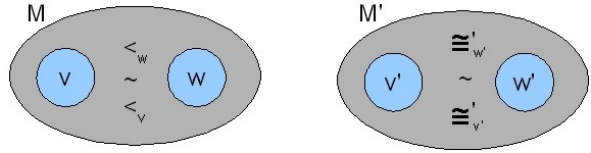

We first prove an auxiliary claim about the right model $\mathbb{M}'$:

**Auxiliary claim.** For all $\varphi \in \mathcal{L}(K, B^c, B^+)$, we have $\mathbb{M}', w' \models \varphi$ iff $\mathbb{M}', v' \models \varphi$.
*Proof of auxiliary claim.* We prove this by induction on the complexity of $\varphi$. The base case, for proposition letters $p$, is true because $V'(p) = \{w', v'\}$ for all $p \in Prop$. The Boolean cases are trivial. We now focus on the three epistemic/doxastic operators:

- $\varphi = K\psi$. Since $[w']_{\sim'} = \{w', v'\} = [v']_{\sim'}$, we have: $\mathbb{M}', w' \models K\psi$ iff ($\mathbb{M}', w' \models \psi$ & $\mathbb{M}', v' \models \psi$) iff $\mathbb{M}', v' \models K\psi$.

- $\varphi = B^+\psi$. Since $[w']_{\sim'} = \{w', v'\} = [v']_{\sim'}$ and $\leq'_{w'} = \{w', v'\} \times \{w', v'\} = \leq'_{v'}$, we have: $\mathbb{M}', w' \models B^+\psi$ iff ($\mathbb{M}', w' \models \psi$ & $\mathbb{M}', v' \models \psi$) iff $\mathbb{M}', v' \models B^+\psi$.

- $\varphi = B^\alpha\psi$. Since $[w']_{\sim'} = [v']_{\sim'}$ and $\leq'_{w'} = \leq'_{v'}$, we have: $\mathbb{M}', w' \models B^\alpha\psi$ iff $\text{Min}_{\leq'_{w'}}(\llbracket\alpha\rrbracket^{\mathbb{M}'} \cap [w']_{\sim'}) \subseteq \llbracket\psi\rrbracket^{\mathbb{M}'}$ iff $\text{Min}_{\leq'_{v'}}(\llbracket\alpha\rrbracket^{\mathbb{M}'} \cap [v']_{\sim'}) \subseteq \llbracket\psi\rrbracket^{\mathbb{M}'}$ iff $\mathbb{M}', v' \models B^\alpha\psi$.

This finishes the proof of the auxiliary claim about $\mathbb{M}'$. We now use this to prove that $\mathbb{M}$ and $\mathbb{M}'$ are statewise $\mathcal{L}(K, B^+, B^c)$-equivalent:

$$\forall \varphi \in \mathcal{L}(K, B^+, B^c) : \big(\mathbb{M}, w \models \varphi \Leftrightarrow \mathbb{M}', w' \models \varphi\big) \text{ and } \big(\mathbb{M}, v \models \varphi \Leftrightarrow \mathbb{M}', v' \models \varphi\big)$$

We prove this again by induction on the complexity of $\varphi$. The base case holds because $V(p) = \{w, v\}$ and $V'(p) = \{w', v'\}$ for all $p \in Prop$. The Boolean cases are trivial. We now focus on the three epistemic/doxastic operators:

- $\varphi = K\psi$

  Since $[w]_\sim = \{w, v\}$ and $[w']_{\sim'} = \{w', v'\}$, we have

  $\mathbb{M}, w \models K\psi \Leftrightarrow (\mathbb{M}, w \models \psi$ and $\mathbb{M}, v \models \psi)$
  $\Leftrightarrow (\mathbb{M}', w' \models \psi$ and $\mathbb{M}', v' \models \psi)$ (induction hypothesis)
  $\Leftrightarrow \mathbb{M}', w' \models K\psi$

  Analogously we prove that $\mathbb{M}, v \models K\psi$ iff $\mathbb{M}', v' \models K\psi$

- $\varphi = B^+\psi$

    Since $\forall x \in W : x \leq_w w \Leftrightarrow x \in \{w, v\}$ and $\forall x' \in W' : x \leq'_{w'} w' \Leftrightarrow x' \in \{w', v'\}$ (and $[w]_\sim = \{w, v\}$ and $[w']_{\sim'} = \{w', v'\}$), we have

    $$\begin{aligned}\mathbb{M}, w \models B^+\psi &\Leftrightarrow (\mathbb{M}, w \models \psi \text{ and } \mathbb{M}, v \models \psi)\\ &\Leftrightarrow (\mathbb{M}', w' \models \psi \text{ and } \mathbb{M}', v' \models \psi) \quad \text{(induction hypothesis)}\\ &\Leftrightarrow \mathbb{M}', w' \models B^+\psi\end{aligned}$$

    Since $\forall x \in W : x \leq_v v \Leftrightarrow x = v$ and $\forall x' \in W' : x \leq'_{v'} v' \Leftrightarrow x' \in \{w', v'\}$ (and $[v]_\sim = \{w, v\}$ and $[v']_{\sim'} = \{w', v'\}$), we have

    $$\begin{aligned}\mathbb{M}, v \models B^+\psi &\Leftrightarrow \mathbb{M}, v \models \psi\\ &\Leftrightarrow \mathbb{M}', v' \models \psi \quad &\text{(induction hypothesis)}\\ &\Leftrightarrow (\mathbb{M}', w' \models \psi \text{ and } \mathbb{M}', v' \models \psi) \quad &\text{(auxiliary claim)}\\ &\Leftrightarrow \mathbb{M}', v' \models B^+\psi\end{aligned}$$

- $\varphi = B^\alpha \psi$

    We make a case distinction:

    **Case A** $v \in [\![\alpha]\!]^\mathbb{M}$

    Then by IH: $\mathbb{M}', v' \models \alpha$. By the auxiliary claim also $\mathbb{M}', w' \models \alpha$, so $[\![\alpha]\!]^{\mathbb{M}'} = \{w', v'\}$. Since $v' \cong'_{w'} w'$ and $[w']_{\sim'} = \{w', v'\}$ we get that $\text{Min}_{\leq'_{w'}}([\![\alpha]\!]^{\mathbb{M}'} \cap [w']_{\sim'}) = \{w', v'\}$. Furthermore, since $\mathbb{M}, v \models \alpha$ and $v <_w w$, we have $\text{Min}_{\leq_w}([\![\alpha]\!]^\mathbb{M} \cap [w]_\sim) = \{v\}$. Hence we have

    $$\begin{aligned}\mathbb{M}, w \models B^\alpha \psi &\Leftrightarrow \text{Min}_{\leq_w}([\![\alpha]\!]^\mathbb{M} \cap [w]_\sim) \subseteq [\![\psi]\!]^\mathbb{M}\\ &\Leftrightarrow \{v\} \subseteq [\![\psi]\!]^\mathbb{M}\\ &\Leftrightarrow \mathbb{M}, v \models \psi\\ &\Leftrightarrow \mathbb{M}', v' \models \psi \quad &\text{(induction hypothesis)}\\ &\Leftrightarrow (\mathbb{M}', w' \models \psi \text{ and } \mathbb{M}', v' \models \psi) \quad &\text{(auxiliary claim)}\\ &\Leftrightarrow \{w', v'\} \subseteq [\![\psi]\!]^{\mathbb{M}'}\\ &\Leftrightarrow \text{Min}_{\leq'_{w'}}([\![\alpha]\!]^{\mathbb{M}'} \cap [w']_{\sim'}) \subseteq [\![\psi]\!]^{\mathbb{M}'}\\ &\Leftrightarrow \mathbb{M}', w' \models B^\alpha \psi\end{aligned}$$

    Analogously we prove that $\mathbb{M}, v \models B^\alpha \psi$ iff $\mathbb{M}', v' \models B^\alpha \psi$.

    **Case B** $v \notin [\![\alpha]\!]^\mathbb{M}$

    Then by IH: $\mathbb{M}', v' \not\models \alpha$. By the auxiliary claim, we get $\mathbb{M}', w' \not\models \alpha$. By IH again, we get $\mathbb{M}, w \not\models \alpha$. Now we have $[\![\alpha]\!]^\mathbb{M} = \emptyset = [\![\alpha]\!]^{\mathbb{M}'}$, and thus

    $$\begin{aligned}\mathbb{M}, w \models B^\alpha \psi &\Leftrightarrow \text{Min}_{\leq_w}([\![\alpha]\!]^\mathbb{M} \cap [w]_\sim) \subseteq [\![\psi]\!]^\mathbb{M}\\ &\Leftrightarrow \text{Min}_{\leq_w}(\emptyset) \subseteq [\![\psi]\!]^\mathbb{M}\\ &\Leftrightarrow \emptyset \subseteq [\![\psi]\!]^\mathbb{M}\\ &\Leftrightarrow \emptyset \subseteq [\![\psi]\!]^{\mathbb{M}'}\\ &\Leftrightarrow \text{Min}_{\leq'_{w'}}(\emptyset) \subseteq [\![\psi]\!]^{\mathbb{M}'}\\ &\Leftrightarrow \text{Min}_{\leq'_{w'}}([\![\alpha]\!]^{\mathbb{M}'} \cap [w']_{\sim'}) \subseteq [\![\psi]\!]^{\mathbb{M}'}\\ &\Leftrightarrow \mathbb{M}', w' \models B^\alpha \psi\end{aligned}$$

    Analogously we prove that $\mathbb{M}, v \models B^\alpha \psi$ iff $\mathbb{M}', v' \models B^\alpha \psi$.

We have now shown that for all $\varphi \in \mathcal{L}(K, B^+, B^c)$, we have $\mathbb{M}, w \models \varphi$ iff $\mathbb{M}', w' \models \varphi$. Yet one easily verifies that $\mathbb{M}, w \models \langle > \rangle \top$, while $\mathbb{M}', w' \not\models \langle > \rangle \top$ (where, of course, $\langle > \rangle = \neg [>] \neg$). Hence, $[>_i]$ is not definable in $\mathcal{L}(K, B^+, B^c)$. □

The [>]-modality is actually so expressive that, together with the knowledge operator, it is able to define the notion of conditional belief — at least, when we restrict ourselves to the *uniform* epistemic plausibility models introduced at the beginning of this subsection.

**Theorem 22.** For all uniform models $\mathbb{M}$, it holds that $\mathbb{M} \models B_i^\alpha \varphi \leftrightarrow K_i((\alpha \wedge \neg \langle >_i \rangle \alpha) \to \varphi)$.

*Proof.* Consider an arbitrary uniform model $\mathbb{M}$ and a state $w$ of $\mathbb{M}$. Note that for any $x \in [w]_{\sim_i}$, it holds that $[w]_{\sim_i} = [x]_{\sim_i}$ (since $\sim_i$ is an equivalence relation) and that $\leq_{i,w} = \leq_{i,x}$ (since $\mathbb{M}$ is uniform). This justifies step ($\ddagger$) below:

$\mathbb{M}, w \models B_i^\alpha \varphi$
$\Leftrightarrow \forall x \in W : x \in \mathrm{Min}_{\leq_{i,w}}(\llbracket \alpha \rrbracket^{\mathbb{M}} \cap [w]_{\sim_i}) \Rightarrow \mathbb{M}, x \models \varphi$
$\Leftrightarrow \forall x \in W : (x \in \llbracket \alpha \rrbracket^{\mathbb{M}} \cap [w]_{\sim_i} \text{ and } \neg \exists y \in \llbracket \alpha \rrbracket^{\mathbb{M}} \cap [w]_{\sim_i} : y <_{i,w} x) \Rightarrow \mathbb{M}, x \models \varphi$ (Fact 19)
$\Leftrightarrow \forall x \in [w]_{\sim_i} : (\mathbb{M}, x \models \alpha \text{ and } \neg \exists y \in [w]_{\sim_i} : (y <_{i,w} x \text{ and } \mathbb{M}, y \models \alpha)) \Rightarrow \mathbb{M}, x \models \varphi$
$\Leftrightarrow \forall x \in [w]_{\sim_i} : (\mathbb{M}, x \models \alpha \text{ and } \neg \exists y \in [x]_{\sim_i} : (y <_{i,x} x \text{ and } \mathbb{M}, y \models \alpha)) \Rightarrow \mathbb{M}, x \models \varphi$ ($\ddagger$)
$\Leftrightarrow \forall x \in [w]_{\sim_i} : (\mathbb{M}, x \models \alpha \text{ and } \mathbb{M}, x \not\models \langle >_i \rangle \alpha) \Rightarrow \mathbb{M}, x \models \varphi$
$\Leftrightarrow \mathbb{M}, w \models K_i((\alpha \wedge \neg \langle >_i \rangle \alpha) \to \varphi)$

$\square$

We now introduce the notion of [>]-bisimilarity, which — as desired — is fully structural:

**Definition 23.** Given epistemic plausiblity models $\mathbb{M} = \langle W, \{\sim_i\}_{i \in G}, \{\leq_{i,w}\}_{i \in G}^{w \in W}, V \rangle$ and $\mathbb{M}' = \langle W', \{\sim_i'\}_{i \in G}, \{\leq_{i,w'}'\}_{i \in G}^{w' \in W'}, V' \rangle$ ; a relation $Z \subseteq W \times W'$ is a [>]-*bisimulation* iff

- if $(w, w') \in Z$, then for all atoms $p$: $w \in V(p) \Leftrightarrow w' \in V'(p)$

- if $(w, w') \in Z$ and $w \sim_i v$ and $v <_{i,w} w$, then there is a $v' \in W'$ such that $(v, v') \in Z$ and $w' \sim_i' v'$ and $v' <_{i,w'}' w'$

- if $(w, w') \in Z$ and $w' \sim_i' v'$ and $v' <_{i,w'}' w'$, then there is a $v \in W$ such that $(v, v') \in Z$ and $w \sim_i v$ and $v <_{i,w} w$

Part 1 of Theorem 24 shows that this is the right notion of bisimulation. Furthermore, we get combined notions of bisimulation in the obvious way. In particular, $\{K, [>]\}$-bisimulations are combined $K$- and [>]-bisimulations; since both of the latter notions are purely structural, also $\{K, [>]\}$-bisimulation is structural. Part 2 of Theorem 24 is the analogue of part 1 for this combined notion. Most importantly, part 3 states that when we restrict ourselves to the class of *uniform* models, we can get equivalence for conditional belief[7] by means of a structural notion of bisimulation. Finally, part 4 says that if we restrict to uniform image-finite models, then (structural) $\{K, [>]\}$-bisimilarity implies $\{K, B^c\}$-bisimilarity (which involves universal quantification over formulas).

**Theorem 24.** Consider two epistemic plausiblity models $\mathbb{M} = \langle W, \{\sim_i\}_{i \in G}, \{\leq_{i,w}\}_{i \in G}^{w \in W}, V \rangle$ and $\mathbb{M}' = \langle W', \{\sim_i'\}_{i \in G}, \{\leq_{i,w'}'\}_{i \in G}^{w' \in W'}, V' \rangle$, and a relation $Z \subseteq W \times W'$.

1. If $Z$ is a [>]-bisimulation, then for all $\varphi \in \mathcal{L}([>])$ and for all $(w, w') \in Z$, it holds that $\mathbb{M}, w \models \varphi \Leftrightarrow \mathbb{M}', w' \models \varphi$.

---

[7]Actually for $\mathcal{L}(K, B^c)$ — but this is no heavy restriction, since it is natural to study both notions simultaneously anyway.

2. If $Z$ is a $\{K, [>]\}$-bisimulation, then for all $\varphi \in \mathcal{L}(K, [>])$ and for all $(w, w') \in Z$, it holds that $\mathbb{M}, w \models \varphi \Leftrightarrow \mathbb{M}', w' \models \varphi$.

3. If $\mathbb{M}$ and $\mathbb{M}'$ are uniform, and $Z$ is a $\{K, [>]\}$-bisimulation, then for all $\varphi \in \mathcal{L}(K, B^c)$ and for all $(w, w') \in Z$, it holds that $\mathbb{M}, w \models \varphi \Leftrightarrow \mathbb{M}', w' \models \varphi$.

4. If $\mathbb{M}$ and $\mathbb{M}'$ are uniform and image-finite, then for any states $w \in W$ and $w' \in W'$, we have that if $w$ and $w'$ are $\{K, [>]\}$-bisimilar, then they are $\{K, B^c\}$-bisimilar as well.

*Proof.* 1. This is easily proved by induction on the complexity of $\varphi$. We only treat the case for $[>_i]$. Suppose that $(w, w') \in Z$ and $\mathbb{M}, w \models [>_i]\varphi$; we show that also $\mathbb{M}', w' \models [>_i]\varphi$. Consider an arbitrary $v' \in W'$ and suppose that $w' \sim_i' v'$ and $v' <'_{i,w'} w'$. By Definition 23, there exists a $v \in W$ such that $(v, v') \in Z$ and $w \sim_i v$ and $v <_{i,w} w$. Since $\mathbb{M}, w \models [>_i]\varphi$ and $w \sim_i v$ and $v <_{i,w} w$, we get $\mathbb{M}, v \models \varphi$. Since $(v, v') \in Z$, we get by the induction hypothesis that also $\mathbb{M}', v' \models \varphi$. The other direction is completely analogous.

2. This follows immediately from part 1 of this theorem and part 1 of Theorem 9.

3. Consider uniform models $\mathbb{M}$ and $\mathbb{M}'$, a $\{K, [>]\}$-bisimulation $Z$, and $(w, w') \in Z$. Now consider an arbitrary $\varphi \in \mathcal{L}(K, B^c)$. Since $\mathbb{M}$ and $\mathbb{M}'$ are uniform, Theorem 22 allows us to systematically delete all occurences of conditional belief operators in $\varphi$, and replace them with $K$- and $[>]$-operators, thus obtaining a formula $\varphi^> \in \mathcal{L}(K, [>])$ such that $\mathbb{M} \models \varphi \leftrightarrow \varphi^>$ (†) and $\mathbb{M}' \models \varphi \leftrightarrow \varphi^>$ (‡).[8] Now we get

$$\begin{aligned}
\mathbb{M}, w \models \varphi &\Leftrightarrow \mathbb{M}, w \models \varphi^> &&(\dagger) \\
&\Leftrightarrow \mathbb{M}', w' \models \varphi^> &&\text{(part 2 of this theorem)} \\
&\Leftrightarrow \mathbb{M}', w' \models \varphi &&(\ddagger)
\end{aligned}$$

4. Assume that $w$ and $w'$ are $\{K, [>]\}$-bisimilar; so there exists a $\{K, [>]\}$-bisimiluation $Z \subseteq W \times W'$ such that $(w, w') \in Z$. Since the models are uniform, we get by part 3 of this theorem that $\mathbb{M}, w \models \varphi \Leftrightarrow \mathbb{M}', w' \models \varphi$ for all $\varphi \in \mathcal{L}(K, B^c)$. Since the models are image-finite, we get by Theorem 13 that $w$ and $w'$ are $\{K, B^c\}$-bisimilar. $\square$

We finish this subsection by providing an overview of the first strategy to solve the main issue of Section 3 (viz. finding a structural notion of bisimulation for conditional belief) and evaluating its advantages and disadvantages.

This strategy has two components. The first component is to impose an extra condition on epistemic plausibility models, viz. uniformity. We argued that this is relatively harmless, since it can be given an intuitive motivation in terms of doxastic/epistemic introspection, and because it is dynamically robust (cf. Theorems 17 and 18). The second component involves what van Benthem calls "redesigning one's language to fit more standard bisimulations" [6, p. 310]. We introduced a new modality $[>]$ and showed that together with knowledge, it can define conditional belief (for uniform models) (cf. Theorems 21 and 22). We then used the structural notion of $\{K, [>]\}$-bisimilarity to establish $\mathcal{L}(K, B^c)$-equivalence and even $\{K, B^c\}$-bisimilarity itself (cf. Theorem 24).

The main disadvantage of this approach lies in its second component: the $[>]$-operator was introduced for the sole purpose of defining conditional belief (while maintaining a struc-tural notion of bisimulation). In itself, however, it does not seem to have any intuitive epistemic/doxastic reading.[9] Therefore, this solution ends up looking a bit *ad hoc*.

---

[8] Making this idea fully formal would require recursively defining a translation function $\tau \colon \mathcal{L}(K, B^c) \to \mathcal{L}(K, [>])$, with as its key clause $\tau(B_i^\alpha \varphi) := K_i((\alpha \wedge \neg \langle >_i \rangle \alpha) \to \varphi)$, and then proving that for uniform models $\mathbb{M}$, we have $\mathbb{M} \models \varphi \leftrightarrow \tau(\varphi)$.

## 4.2 Assuming connectedness

The second approach tries to keep the advantages of the first one, while avoiding its major drawback, viz. the ad hoc introduction of new operators. The basic idea is that, with an extra condition on the epistemic plausibility models, conditional belief can be reduced to knowledge and safe belief. Hence, the $B^+$-operator plays the role of the $[>]$-operator in the previous approach, but unlike the $[>]$-operator, it *does* have an intuitive doxastic interpretation. The extra condition on the models that we need is local connectedness:

**Definition 25.** An epistemic plausibility model $\mathbb{M} = \langle W, \{\sim_i\}_{i \in G}, \{\leq_{i,w}\}_{i \in G}^{w \in W}, V \rangle$ is called *locally connected* iff for all agents $i \in G$ and states $w, v \in W$ it holds that if $w \sim_i v$, then $w \leq_{i,w} v$ or $v \leq_{i,w} w$.

Whether this is a natural condition is a bit more doubtful than in the case of uniformity. At least, local connectedness is dynamically robust:

**Theorem 26.** If an epistemic plausibility model $\mathbb{M}$ is locally connected, then so are $\mathbb{M}!\varphi$ and $\mathbb{M} \Uparrow \varphi$.

*Proof.* Consider a locally connected model $\mathbb{M}$. We first treat the case of public announcement of $\varphi$. Consider arbitrary states $w, v$ of $\mathbb{M}!\varphi$ and suppose that $w \sim_i^{!\varphi} v$. By Definition 3, we get that $w \sim_i v$. Since $\mathbb{M}$ is locally connected, it follows that $w \leq_{i,w} v$ or $v \leq_{i,w} w$. Since $\leq_{i,w}^{!\varphi} = \leq_{i,w}$ (cf. Definition 3), we get that also $w \leq_{i,w}^{!\varphi} v$ or $v \leq_{i,w}^{!\varphi} w$, as desired.

We now treat the case of radical upgrade with $\varphi$. Consider arbitrary states $w, v$ of $\mathbb{M} \Uparrow \varphi$ and suppose that $w \sim_i^{\Uparrow\varphi} v$. By Definition 3, we get that $w \sim_i v$. Since $\mathbb{M}$ is locally connected, it follows that $w \leq_{i,w} v$ or $v \leq_{i,w} w$. We now make the following case distinction:

- $\mathbb{M}, w \models \varphi$, $\mathbb{M}, v \models \varphi$. Then by Definition 3, $w \leq_{i,w}^{\Uparrow\varphi} v$ iff $w \leq_{i,w} v$, and $v \leq_{i,w}^{\Uparrow\varphi} w$ iff $v \leq_{i,w} w$. Since $w \leq_{i,w} v$ or $v \leq_{i,w} w$, it follows that $w \leq_{i,w}^{\Uparrow\varphi} v$ or $v \leq_{i,w}^{\Uparrow\varphi} w$, as required.

- $\mathbb{M}, w \not\models \varphi$, $\mathbb{M}, v \not\models \varphi$. Analogous to the previous case.

- $\mathbb{M}, w \models \varphi$, $\mathbb{M}, v \not\models \varphi$. Then by Definition 3, $w \leq_{i,w}^{\Uparrow\varphi} v$.

- $\mathbb{M}, w \not\models \varphi$, $\mathbb{M}, v \models \varphi$. Then by Definition 3, $v \leq_{i,w}^{\Uparrow\varphi} w$.

□

We now show that, when we require the models to be both uniform (cf. the previous subsection) and locally connected, then conditional belief can be defined in terms of knowledge and safe belief.[10]

**Theorem 27.** For all uniform and locally connected models $\mathbb{M}$, it holds that $\mathbb{M} \models B_i^\alpha \varphi \leftrightarrow (\hat{K}_i \alpha \to \hat{K}_i(\alpha \land B_i^+(\alpha \to \varphi)))$.

---

[9] Cf. "The intuitive meaning of these operators [such as $[>]$, *LD*] is not very clear, but they can be used to define other interesting modalities, capturing various 'doxastic attitudes'.", [1, p. 32].

[10] A similar definition was already proposed in the context of modal conditional logics of normality; cf. Boutilier[3, p. 104].

*Proof.* Consider an arbitrary uniform and locally connected model $\mathbb{M}$ and a state $w$ of $\mathbb{M}$. We will only work in the model $\mathbb{M}$, and can therefore write $x \models \psi$ instead of $\mathbb{M}, x \models \psi$. We prove both directions:

- $w \models B_i^\alpha \varphi \to \left( \hat{K}_i \alpha \to \hat{K}_i(\alpha \wedge B_i^+(\alpha \to \varphi)) \right)$

  Assume $w \models B_i^\alpha \varphi$ and $w \models \hat{K}_i \alpha$; we show that $w \models \hat{K}_i(\alpha \wedge B_i^+(\alpha \to \varphi))$. Since $w \models \hat{K}_i \alpha$, we have that $[\![\alpha]\!]^\mathbb{M} \cap [w]_{\sim_i} \neq \emptyset$. By well-foundedness of $\leq_{i,w}$, also $\mathrm{Min}_{\leq_{i,w}}([\![\alpha]\!]^\mathbb{M} \cap [w]_{\sim_i}) \neq \emptyset$. Therefore we can pick a state $x \in \mathrm{Min}_{\leq_{i,w}}([\![\alpha]\!]^\mathbb{M} \cap [w]_{\sim_i})$. Note that $w \sim_i x$ and $x \models \alpha$. Hence, if we can show that also $x \models B_i^+(\alpha \to \varphi)$, then we get $w \models \hat{K}_i(\alpha \wedge B_i^+(\alpha \to \varphi))$, as required.

  We now prove that $x \models B_i^+(\alpha \to \varphi)$. Consider an arbitrary $y \in [x]_{\sim_i}$ and suppose that $y \leq_{i,x} x$ and $y \models \alpha$; we will show that $y \models \varphi$. Since $w \models B_i^\alpha \varphi$, it suffices to show that $y \in \mathrm{Min}_{\leq_{i,w}}([\![\alpha]\!]^\mathbb{M} \cap [w]_{\sim_i})$.

  We already know that $y \models \alpha$ and since $y \sim_i x$ and $x \sim_i w$, also $y \sim_i w$; hence $y \in [\![\alpha]\!]^\mathbb{M} \cap [w]_{\sim_i}$. We now establish the $\leq_{i,w}$-minimality of $y$ in $[\![\alpha]\!]^\mathbb{M} \cap [w]_{\sim_i}$. Consider an arbitrary $z \in [\![\alpha]\!]^\mathbb{M} \cap [w]_{\sim_i}$ and suppose that $z \leq_{i,w} y$; we show that $y \leq_{i,w} z$. Since $y \leq_{i,x} x$ and $x \sim_i w$, we get by uniformity that $y \leq_{i,w} x$. Together with $z \leq_{i,w} y$, we get $z \leq_{i,w} x$. Since $z \in [\![\alpha]\!]^\mathbb{M} \cap [w]_{\sim_i}$ and $x$ is $\leq_{i,w}$-minimal in $[\![\alpha]\!]^\mathbb{M} \cap [w]_{\sim_i}$, it follows that $x \leq_{i,w} z$. Together with $y \leq_{i,w} x$, this implies that $y \leq_{i,w} z$.

- $w \models \left( \hat{K}_i \alpha \to \hat{K}_i(\alpha \wedge B_i^+(\alpha \to \varphi)) \right) \to B_i^\alpha \varphi$

  Assume $w \models \hat{K}_i \alpha \to \hat{K}_i(\alpha \wedge B_i^+(\alpha \to \varphi))$ $(*)$; we show that $w \models B_i^\alpha \varphi$. Consider an arbitrary $x \in \mathrm{Min}_{\leq_{i,w}}([\![\alpha]\!]^\mathbb{M} \cap [w]_{\sim_i})$; we will show that $x \models \varphi$. Note that since $x \in [\![\alpha]\!]^\mathbb{M} \cap [w]_{\sim_i}$, we have $w \models \hat{K}_i \alpha$, and thus by $(*)$ we get $w \models \hat{K}_i(\alpha \wedge B_i^+(\alpha \to \varphi))$. Hence there is a state $y \in [w]_{\sim_i}$ such that $y \models \alpha$ and $y \models B_i^+(\alpha \to \varphi)$.

  We now show that $x \leq_{i,y} y$. Since $x \sim_i w$ and $y \sim_i w$, we get that $x \sim_i y$. Since $\mathbb{M}$ is locally connected, it follows that $x \leq_{i,y} y$ or $y \leq_{i,y} x$. In the first case, we're done. Now suppose that the second case obtains. Since $y \sim_i w$, we get by uniformity that $y \leq_{i,w} x$. Since $y \in [\![\alpha]\!]^\mathbb{M} \cap [w]_{\sim_i}$ and $x$ is $\leq_{i,w}$-minimal in $[\![\alpha]\!]^\mathbb{M} \cap [w]_{\sim_i}$, it follows that $x \leq_{i,w} y$. Again by uniformity, we get $x \leq_{i,y} y$.

  We have proved that $x \leq_{i,y} y$. Also recall that $x \sim_i y$. Hence, it follows from $y \models B_i^+(\alpha \to \varphi)$ that $x \models \alpha \to \varphi$. Since $x \models \alpha$ we get that $x \models \varphi$.

$\square$

Using this definability result, we can now immediately prove the analogon of Theorem 24; of course, since we did not have to introduce a new modality and a new notion of bisimulation corresponding to it, we only reformulate its third and fourth part. The importance of this is that when we restrict ourselves to the class of uniform and locally connected models, we can get equivalence for conditional belief by means of a structural notion of bisimulation, viz. $\{K, B^+\}$-bisimulation. Furthermore, if we restrict to the uniform, locally connected *and* image-finite models, then (structural) $\{K, B^+\}$-bisimulation implies $\{K, B^c\}$-bisimilarity (which involves universal quantification over formulas).

**Theorem 28.** Consider two epistemic plausiblity models $\mathbb{M} = \langle W, \{\sim_i\}_{i \in G}, \{\leq_{i,w}\}_{i \in G}^{w \in W}, V \rangle$ and $\mathbb{M}' = \langle W', \{\sim_i'\}_{i \in G}, \{\leq_{i,w'}'\}_{i \in G}^{w' \in W'}, V' \rangle$, and a relation $Z \subseteq W \times W'$.

1. If $\mathbb{M}$ and $\mathbb{M}'$ are uniform and locally connected, and $Z$ is a $\{K, B^+\}$-bisimulation, then for all $\varphi \in \mathcal{L}(K, B^+, B^c)$ and for all $(w, w') \in Z$, it holds that $\mathbb{M}, w \models \varphi \Leftrightarrow \mathbb{M}', w' \models \varphi$.

2. If $\mathbb{M}$ and $\mathbb{M}'$ are uniform, locally connected and image-finite, then for any states $w \in W$ and $w' \in W'$, we have that if $w$ and $w'$ are $\{K, B^+\}$-bisimilar, then they are $\{K, B^c\}$-bisimilar as well.

*Proof.* 1. Consider uniform and locally connected models $\mathbb{M}$ and $\mathbb{M}'$, a $\{K, B^+\}$-bisimulation $Z$, and $(w, w') \in Z$. Now consider an arbitrary $\varphi \in \mathcal{L}(K, B^+, B^c)$. Since $\mathbb{M}$ and $\mathbb{M}'$ are uniform and locally connected, Theorem 27 allows us to systematically delete all occurences of conditional belief operators in $\varphi$, and replace them with $K$- and $B^+$-operators, thus obtaining a formula $\varphi^+ \in \mathcal{L}(K, B^+)$ such that $\mathbb{M} \models \varphi \leftrightarrow \varphi^+$ (†) and $\mathbb{M}' \models \varphi \leftrightarrow \varphi^+$ (‡). Now we get

$$
\begin{aligned}
\mathbb{M}, w \models \varphi &\Leftrightarrow \mathbb{M}, w \models \varphi^+ & (\dagger) \\
&\Leftrightarrow \mathbb{M}', w' \models \varphi^+ & \text{(second part of Theorem 11)} \\
&\Leftrightarrow \mathbb{M}', w' \models \varphi & (\ddagger)
\end{aligned}
$$

2. Assume that $w$ and $w'$ are $\{K, B^+\}$-bisimilar; so there exists a $\{K, B^+\}$-bisimiluation $Z \subseteq W \times W'$ such that $(w, w') \in Z$. Since the models are uniform and locally connected, we get by part 1 of this theorem that $\mathbb{M}, w \models \varphi \Leftrightarrow \mathbb{M}', w' \models \varphi$ for all $\varphi \in \mathcal{L}(K, B^c)$. Since the models are image-finite, we get by Theorem 13 that $w$ and $w'$ are $\{K, B^c\}$-bisimilar. □

Just as we did in the previous subsection, we now provide an overview of the second strategy to solve the main issue of Section 3. This approach reduced conditional belief to knowledge and safe belief, which are both intuitively clear epistemic/doxastic notions. Therefore, the main issue of the first approach, viz. the *ad hoc* character of its introduction of the $[>]$-operator, is avoided. In order to get the desired results about $\mathcal{L}(K, B^c)$-equivalence and $\{K, B^c\}$-bisimilarity (cf. Theorem 28), we required the epistemic plausibility models to be not only uniform, but also locally connected. The uniformity constraint inherits of course all of its justifications (intuitive epistemic/doxastic interpretation and dynamic robustness) from the previous subsection. However, the new constraint, local connectedness, seems to be less motivated: while it is also dynamically robust (cf. Theorem 26), it might not have as intuitive an interpretation as the uniformity constraint.

## 5 Dynamics and bisimulation

In this section, we will make some remarks about the interaction between bisimulation and dynamic model changes. First, however, we need to decide which approach to conditional belief is to be adopted. Throughout this paper, we proposed three approaches: a non-structural one in Section 3, and two structural ones in Section 4. We already argued extensively that the structural approaches are to be preferred over the non-structural one. Of the two structural approaches, however, none seemed to be highly preferable over the other. For reasons that will become clear later, we will henceforth adopt the approach developed in Subsection 4.2. However, one should keep in mind that this section could easily be rewritten in terms of the approach developed in Subsection 4.1.

The main use of bisimulations is to prove $\mathcal{L}$-equivalence of two models (for some language $\mathcal{L}$). Fact 5 tells us that adding the dynamic operators $[!\varphi]$ and $[\Uparrow \varphi]$ to the language $\mathcal{L}(K, B^+, B^c)$ does not increase its expressivity: using the reduction axioms, every formula of the dynamic language $\mathcal{L}(K, B^+, B^c, !, \Uparrow)$ can be rewritten as an equivalent formula of the original static language $\mathcal{L}(K, B^+, B^c)$. Thus, information about what will be the case after some change has taken place can be *pre-encoded* in the static language. We will now combine this pre-encoding strategy with Theorem 28:

**Theorem 29.** Consider two uniform and locally connected epistemic plausiblity models $\mathbb{M} = \langle W, \{\sim_i\}_{i \in G}, \{\leq_{i,w}\}_{i \in G}^{w \in W}, V \rangle$ and $\mathbb{M}' = \langle W', \{\sim'_i\}_{i \in G}, \{\leq'_{i,w'}\}_{i \in G}^{w' \in W'}, V' \rangle$, states $w \in W$ and $w' \in W'$, and a $\{K, B^+\}$-bisimulation $Z \subseteq W \times W'$ such that $(w, w') \in Z$. Furthermore, consider an arbitrary formula $\varphi \in \mathcal{L}(K, B^+, B^c)$; then:

1. If $\mathbb{M}, w \models \varphi$ and $\mathbb{M}', w' \models \varphi$, then $\forall \psi \in \mathcal{L}(K, B^+, B^c) : \mathbb{M}!\varphi, w \models \psi \Leftrightarrow \mathbb{M}'!\varphi, w' \models \psi$.

2. $\forall \psi \in \mathcal{L}(K, B^+, B^c) : \mathbb{M} \Uparrow \varphi, w \models \psi \Leftrightarrow \mathbb{M}' \Uparrow \varphi, w' \models \psi$.

*Proof.* 1. Consider an arbitrary $\psi \in \mathcal{L}(K, B^+, B^c)$; hence $[!\varphi]\psi \in \mathcal{L}(K, B^+, B^c, !, \Uparrow)$. Using Fact 5, we can find a formula $([!\varphi]\psi)^s \in \mathcal{L}(K, B^+, B^c)$ that is equivalent to $\psi$. We now have:

$$\begin{aligned}
\mathbb{M}!\varphi, w \models \psi &\Leftrightarrow \mathbb{M}, w \models [!\varphi]\psi &&\text{(since } \mathbb{M}, w \models \varphi\text{)} \\
&\Leftrightarrow \mathbb{M}, w \models ([!\varphi]\psi)^s \\
&\Leftrightarrow \mathbb{M}', w' \models ([!\varphi]\psi)^s &&\text{(part 1 of Theorem 28)} \\
&\Leftrightarrow \mathbb{M}', w' \models [!\varphi]\psi \\
&\Leftrightarrow \mathbb{M}'!\varphi, w' \models \psi &&\text{(since } \mathbb{M}', w' \models \varphi\text{)}
\end{aligned}$$

The proof of 2. is analogous (or even easier, as we don't have to worry about preconditions anymore) □

We finish by making two remarks. We proved the previous theorem by combining the pre-encoding strategy and Theorem 28. Hence, if we would have adopted the first approach to bisimulations for conditional belief (i.e. that developed in Subsection 4.1), we would have needed reduction axioms for $[!\varphi][>_i]\psi$ and $[\Uparrow \varphi][>_i]\psi$. These turn out to be almost identical to those for safe belief.

**Fact 30.** The following are sound with respect to epistemic plausibility models:

$$\begin{aligned}
[!\varphi][>_i]\psi &\leftrightarrow (\varphi \to [>_i][!\varphi]\psi) \\
[\Uparrow \varphi][>_i]\psi &\leftrightarrow \big(\varphi \to [>_i](\varphi \to [\Uparrow \varphi]\psi)\big) \land \\
&\qquad \big(\neg\varphi \to ([>_i](\neg\varphi \to [\Uparrow \varphi]\psi) \land K_i(\varphi \to [\Uparrow \varphi]\psi))\big)
\end{aligned}$$

The second remark is about the *strength* of bisimulation. Theorem 28 tells us that bisimulation 'now' implies modal equivalence 'now'. Theorem 29, however, tells us that bisimulation 'now' implies modal equivalence '*later*' (i.e. after the model has undergone some dynamic effects). Since both uniformity and local connectedness are dynamically robust (cf. Theorems 18 and 26), Theorem 29 can be repeated to prove that the same holds for any sequence of epistemic dynamics (e.g. if $\mathbb{M}, w$ and $\mathbb{M}', w'$ are $\{K, B^+\}$-bisimilar, then $(((\mathbb{M}!\varphi_1) \Uparrow \varphi_2) \Uparrow \varphi_3)!\varphi_4, w$ and $(((\mathbb{M}'!\varphi_1) \Uparrow \varphi_2) \Uparrow \varphi_3)!\varphi_4, w'$ are $\mathcal{L}(K, B^+, B^c)$-equivalent — provided they survive the public announcements, of course). Hence, if two epistemic plausibility models are bisimilar at one point, then their entire epistemic-doxastic futures are indistinguishable.

# 6 Conclusion

The aim of this paper has been to explore the model theory of epistemic plausibility models, which has been largely ignored in the present literature. We focused on the notion of bisimulation, and proved various bisimulation-implies-equivalence type theorems, a Hennesy-Milner type theorem, and two undefinability results. However, our main conclusion is a negative one, viz. that bisimulations cannot straightforwardly be generalized to epistemic plausibility models if conditional belief is taken into account. We presented and compared two different ways of coping with this issue: adding a modality to the language, and putting extra constraints on the models. Finally, we established some results about the interaction between bisimulation and dynamic model changes, and commented on the strength of bisimulation to establish equivalence 'now and in the future'.

As this is one of the first papers on the model theory of epistemic plausibility models, there is obviously still much work to be done in this area. The main question asked in this paper, viz. what is the right notion of bisimulation for conditional belief, has not yet received a fully satisfactory answer. More work is needed on comparing the approaches developed in this paper, but also on developing still other approaches that have thus far remained under our radar. Finally, this paper has focused almost exclusively on the topic of bisimulation, but one can also consider other topics from the model theory of modal logic, such as relations with first-order logic (via the standard translation), and investigate whether/how they can be generalized.